\documentclass[twocolumn,showpacs,amsmath,amssymb,floatfix,pre]{revtex4-1}
\usepackage{epsfig}
\usepackage{graphicx}
\usepackage{subfigure}
\usepackage{float}
\usepackage{multirow}
\usepackage{amssymb}
\usepackage{amsmath}
\usepackage{picinpar}

\usepackage{ae,dsfont}

\newcommand{\be}{\begin{equation}}
\newcommand{\ee}{\end{equation}}
\newcommand{\ba}{\begin{eqnarray}}
\newcommand{\ea}{\end{eqnarray}}
\newcommand{\para}{/\hspace{-0.08cm}/}

\begin{document}

\title{Mean-field calculation of critical parameters and log-periodic characterization
of an aperiodic-modulated model}

\author{T. P. Oliveira} 
\email{tharnier@fisica.ufsc.br}
\author{N. S. Branco} 
\email{nsbranco@fisica.ufsc.br}
\affiliation{Departamento de Física,
Universidade Federal de Santa Catarina, 
88040-900, Florianópolis, SC, Brazil} 

\date{\today}

\begin{abstract}

        We employ a mean-field approximation to study the Ising model with aperiodic modulation
of its interactions in one spatial direction. Two different values for the exchange constant, $J_A$ and
$J_B$, are present, according to the Fibonacci sequence. We calculated the pseudo-critical temperatures 
for finite systems and extrapolate them to the thermodynamic limit. We explicitly obtain the exponents 
$\beta$, $\delta$, and $\gamma$ and, from the usual scaling relations for anisotropic models
at the upper critical dimension (assumed to be 4 for the model we treat), we calculate 
$\alpha$, $\nu$, $\nu_{\para}$, $\eta$, and $\eta_{\para}$. Within the framework of a
renormalization-group approach, the Fibonacci sequence is a marginal one and we obtain exponents
which depend on the ratio $r \equiv J_B/J_A$, as expected. But the scaling relation
$\gamma = \beta \left( \delta -1  \right)$ is obeyed for all values of $r$ we studied.
We characterize some thermodynamic functions as log-periodic functions of their
arguments, as expected for aperiodic-modulated models, and obtain precise values for
the exponents from this characterization.

\noindent     

\end{abstract}
\pacs{05.50.+q, 64.60.F-, 64.60.-i}

\maketitle

\newpage

\section{Introduction} \label{sec:introduction}

      Non-uniform systems are interesting and important both from the theoretical and
experimental points of view. Experimentally, there are already several techniques of surface growth 
\cite{ross,shchukin,williams} that let one controls the layout of the layers in order to follow, for
example, an aperiodic sequence. On the other hand, many theoretical issues may be
raised, concerning the behavior of systems with random disorder or aperiodic modulations
of their interactions; it is the last case that concerns us in this work. More specifically, our interest
is to calculate the critical parameters of the Ising model within a mean-field framework and
to characterize the log-periodic behavior of some thermodynamic quantities. 

        The interactions of the  model we treat can assume one between two different values, and 
are ordered according to the Fibonacci aperiodic sequence. For models that have a continuous
transition in its uniform version, the influence of aperiodic modulations
on their critical behavior is determined by the Harris-Luck criterion \cite{luck1} (which seems
to hold true for models with first-order transition as well \cite{girardi2011}).
According to this criterion, the Fibonacci sequence is a marginal one;
several results show that a marginal perturbation leads to a dependence of the
critical exponents on the ratio between the two different interactions \cite{igloi,turban,faria}. 
Using the simplest version of a mean-field approximation, we confirm these results and expand
them to include other critical exponents, in order to test scaling relations, and to characterize
log-periodic oscillations.

   The rest of this work is organized as follows. In the next section we present some properties of
aperiodic sequences, define the model we treat, and outline the mean-field approximation we use.
Our results are shown and discussed in Section \ref{sec:results} and in Section \ref{sec:conclusion}
we summarize our findings.

   \section{Aperiodic sequences and mean-field approximation}\label{sec:aperiodic}
    
    Aperiodic sequences may be used, for example, to model quasicrystals \cite{shechtman}: 
interactions vary according to the order embodied in the sequences.
These are built from substitution rules, in such a way that no subset of the sequence
is ever repeated. In our case, we define an Ising model on a hypercubic lattice,
of coordination number $z$, given by the Hamiltonian:
\be
     {\cal H} = - \sum_{<i,j>} J_{ij} S_i S_j,    \label{eq:hamiltoniana}
\ee
such that the sum is over nearest-neighbor pairs on the lattice, $J_{ij}$ is the exchange constant
between spins $S_i$ and $S_j$, which can assume the values $J_A$ and $J_B$ in a particular
spatial direction,
according to the respective letter in the aperiodic sequence, and $S_i = \pm 1$, $\forall i$. 

  In this work we are particularly interested in the Fibonacci sequence, which is obtained
from the substitution rules: 
 \be
          A  \rightarrow s(A) = AB,\;\;  B  \rightarrow s(B)= A. \label{eq:sequence}
     \ee
This means that, from one stage of the construction of the aperiodic sequence to the next,
all letters $A$ are replaced by $AB$ and all letters $B$ are replaced by $A$. Starting with
the letter $A$, the first stages of this sequence are: 
$A \rightarrow AB \rightarrow ABA \rightarrow
ABAAB \rightarrow ABAABABA $. This last finite sequence corresponds to the following
sequence of interaction constants: $J_A J_B J_A J_A J_B J_A J_B J_A$. In one of the
spatial dimension of the hypercubic lattice (horizontal, say) the exchange constants 
follow this sequence, while in
the remaining perpendicular hypersurface all interactions assume the same value, 
which is the same as for the succeeding horizontal bonds.
An example of a lattice constructed
this way, in two dimensions, is depicted in Fig. \ref{fig:rede}.

   One of the interesting theoretical questions one may pose is about the influence of
aperiodic modulations on the critical behavior of the model, when compared
to its uniform counterpart. For the case of continuous transition on the uniform model,
the Harris-Luck criterion determines whether the introduction of a given aperiodic modulation
changes the universality class or not \cite{luck1}. This
change is determined by the crossover exponent $\Phi$, given by:
\be
          \Phi = 1 + d_a \nu (\omega - 1),    \label{eq:Phi}
\ee
where $\omega$ is the exponent describing the behavior of geometrical fluctuations of the sequence 
(see below), $d_a$ is the number 
of dimensions upon which the aperiodic sequence acts ($d_a=1$ in our case) 
and $\nu$ is the correlation-length critical exponent 
of the uniform model. When $\Phi>0$ the introduction of the aperiodic sequence changes the
critical exponents from the values assumed for the uniform model (the sequence is said relevant in
this case) and when $\Phi<0 $ the critical behavior of the aperiodic model is the same
as for the uniform one (irrelevant sequence). For $\Phi=0$, the sequence is marginal and 
previous results show that the
critical exponents are non-universal: they depend on the ratio $r \equiv J_B/J_A$ \cite{igloi}.
In the mean-field framework, $\nu=1/2$ for the uniform Ising model and the crossover exponent
reduces to (see Eq. \ref{eq:Phi}):
\be 
     \Phi = \frac{1}{2} ( 1 + \omega ). \label{eq:campomedio}
\ee

    Therefore, for $\omega=-1$ the sequence is marginal. which is the case for the Fibonacci
sequence, as we will shortly see. This quantity and others properties of two-letter
sequences are obtained from
their substitution matrix ${\cal M}$, which is defined as:
     \be \label{eq:matrizsubs}
          {\cal M}=\begin{pmatrix} n_i^{s(i)} & n_i^{s(j)}  
          \\                    n_j^{s(i) } & n_j^{s(j)}\end{pmatrix},
     \ee
where $n_i^{s(j)}$ is the number of letters $i$ that are generated by
applying the rule $s(j)$. Several features of the sequences are determined by
the eigenvalues of ${\cal M}$. The greatest eigenvalue ($\lambda_1$) determines the rate of growth of 
the total number of letters ${\cal N}$, such that ${\cal N}\sim\lambda_1^n$, $n\gg1$, where $n$ is the
number of iterations in the construction of the sequence.
The second greatest eigenvalue ($\lambda_2$) 
determines the {\it wandering  exponent} $\omega$ (Eqs. \ref{eq:Phi} and \ref{eq:campomedio}) 
through:
     \be
          \omega=\frac{\ln |\lambda_2|}{\ln \lambda_1},
     \ee
such that the fluctuation in one of the letters, $g$, is given by \cite{faria}:
     \be
          g\sim {\cal N}^{\omega}.
     \ee

    For the Fibonacci sequence:
        \be \label{eq:Fibonaccimatriz}
          {\cal M}=\begin{pmatrix} 1 & 1 
          \\                    1 & 0 \end{pmatrix},
     \ee
and $\lambda_1 = (1 + \sqrt{5})/2$ and $\lambda_2=-\lambda_1^{-1}$, such that $\omega=-1$,
as anticipated.  

   Therefore, the aperiodic modulation obtained with the Fibonacci sequence, within the mean-field
approximation applied to the Ising model, is a marginal one, when the sequence acts
on one of the spatial directions.

     We study the present model (Eq. \ref{eq:hamiltoniana}) within the simplest mean-field approximation. 
It may be obtained either from
the Bogoliubov inequality \cite{callen} with a single-spin trial Hamiltonian or, in a less rigorous
framework, from substituting the
magnetization $m_i$ for the spin $S_i$. Due to the aperiodic modulation, the values of $m_i$ vary along
the direction upon which the aperiodic sequence acts (although they are the same for a given hyperplane 
perpendicular to this direction). The system of equations one has to solve is:
\ba
   m_i & = & \tanh \left[ K_{i-1} m_{i-1} + (z-2) K_i m_i + K_i m_{i+1} + h \right], \nonumber \\
           &    &  i = 1,..,N \label{eq:sistema}
\ea
where $K_l \equiv \beta J_l$, $h \equiv \beta H$, $\beta \equiv 1/k_B T$, $H$ is a uniform
magnetic field, $k_B$ is Boltzmann constant, $T$ is the temperature, and $N$ is the number of
hyperplanes on the system (or, equivalently, the size of the aperiodic sequence). 
       
\section{Results} \label{sec:results}

      \subsection{Critical temperatures}
 
     The first task is to obtain the critical temperature $T_c$; our strategy is to calculate
pseudo-critical temperatures for finite systems and extrapolate the data to the thermodynamic
limit. Since the transition is expected to be a continuous one and our goal is to calculate
$T_c$, we can expand Eqs. \ref{eq:sistema}
with $H=0$ up to first order on the magnetizations:
\be
   \mathds{K} . \vec{m} = 0,
\ee
where 
\be 
    \mathds{K}  = \label{eq:matriz}
                        \begin{pmatrix} \tilde{K_1} & K_1 & 0 & 0 & ... & 0 \\
                                                      K_1 & \tilde{K_2} & K_2 & 0 & ... & 0 \\
                                                      0 & K_2 & \tilde{K_3} & K_3 & ... & 0 \\
                                                         &     &   &  \vdots &   &  \\
                                                      0 & 0 &  0 & ... & K_{N-1} & \tilde{K_N} \end{pmatrix}
\ee
and
\be
    \vec{m}  =  \left(  \begin{array}{c}
                                 m_1 \\
                                 m_2 \\
                                 m_3 \\
                                 \vdots  \\
                                  m_N \end{array} \right)   \label{eq:vetorm},
\ee
with
$\tilde{K_i} \equiv (z-2) K_i -1$ and free boundary conditions ($m_{N+1}=m_0=0$).
The interaction parameters $K_1, K_2, ... , K_N$ assume
the values $K_A$ or $K_B$ according to the respective letter on the Fibonacci sequence.
Since we do not expect the critical exponents to depend on $z$, we have worked only with 
$z=6$ to obtain the critical parameters.

   For temperatures greater than the pseudo-critical one, the only solution to this system is
$\vec{m}=0$. So, the matrix $ \mathds{K}$ has an inverse, i.e.,  $\det(\mathds{K}) \neq 0$ 
for this region of temperatures. Therefore, coming from above, the first temperature such that
$\det(\mathds{K}) = 0$ is the pseudo-critical temperature. This procedure is applied to systems
with different linear sizes $L$ (corresponding to the length of the aperiodic sequence, $N$)
and extrapolated to $L \rightarrow \infty$. One has to be sure that
the first temperature such that $\det(\mathds{K}) = 0$ is actually obtained, since many temperatures
satisfy this criterion below the first one and they tend to accumulate close to the pseudo-critical
temperature as $L$ increases.

In order to extrapolate our results to the thermodynamic limit, we have used the so-called $BST$
extrapolation \cite{bst} in two different ways (see below). The errors of our evaluations are obtained as usual
for the this method of extrapolation \cite{bst}. 

    Since we expect log-periodic oscillations on models 
with aperiodic-modulated interactions, the pseudo-critical
temperatures do not converge monotonically to the thermodynamic values: on top of an apparent overall
convergence, there are oscillations on the values for finite $L$. Therefore, we have also applied the 
$BST$ procedure
to every other value of the pseudo-critical temperatures. Both procedures lead to the same values in
the thermodynamic limit. In Table \ref{table:Tc} we show our results for the critical temperatures for
some values of the ratio $r$, extrapolated from pseudo-critical temperatures obtained for $L$ up
to $121,393$ for $r=0.5$ and $1.3$ and up to $196,418$ for the other values of $r$.
Note that we show ten decimal figures for $r \neq 1$, which is certainly enough
to obtain precise values for the critical exponents. For $r=1$ we show all figures we are able to obtain, 
since we can compare it to the expected value within the mean-field approximation: there is an agreement
up to 15 decimal figures. 

\begin{table}
  \centering
\begin{tabular}{| c | c |} \hline
  $r$  & $k_B T_c/J_A$ \\ \hline
  0.5 & 5.2939768858 \\
  0.7 & 5.4801586902 \\
  1.0 & 6.000000000000038(64) \\
  1.3 & 6.8300746634 \\
  1.5 & 7.4992699398 \\ \hline
\end{tabular}
 \caption{Extrapolated critical temperatures for some values of
the ratio $r \equiv J_B/J_A$. For $r=1$ (uniform model) we obtain, within error bars,
the exact value, $k_B T_c/J=z=6$.}\label{table:Tc}
\end{table}
     
    In Fig. \ref{fig:comparacaoPalagyi} we compare our values for $T_c$ with those obtained in Ref. 
\onlinecite{igloi}. The quantity $T_c^0$ is the critical temperature for a uniform model with the same
{\it mean} value $\bar{J}$ for the interaction constant $J$  for a given $r$. More precisely,
$\bar{J} \equiv J_A (p_A + r p_B)$, where $p_A$ and $p_B$ are the fraction of letters $A$ and $B$, respectively, on the
infinite aperiodic sequence. These fractions are obtained from the entries of the eigenvector corresponding to
the greatest eigenvalue of the substitution matrix. We notice the agreement is quite good; the apparent
difference for some regions of $r$ comes from the fact that we have few data points and have made an interpolation of our data.

   \subsection{Magnetization}

     Having calculated the critical temperatures, we can now obtain, from the original system of equations
(Eq. \ref{eq:sistema}), the magnetization for each plane. The goal is to solve this system for $m_i$ for different values
of the reduced temperature $t \equiv (T-T_c)/T_c$ and of the reduced magnetic field $h$ ($\equiv \beta H$).
In order to accomplish this  we have
tested three procedures: the first one based on the Newton method \cite{numericalrecipes}, the second one uses
the secant method \cite{numericalrecipes} and finally the so-called fixed-point method \cite{lindfield}. We analyzed the
convergence time, for large systems and for small values of the reduced temperature, and the accuracy (with respect to known results for 
small lattices). The first method is the less precise, the secant method is the most efficient for small
values of $t$, and the fixed-point method is the most efficient for large lattices. We have chosen the last one, to be able to go to
larger systems.

     After a predetermined accuracy is achieved, within the fixed-point method, we stop the iterations  and calculate the
mean magnetization as:
\be
    m(L) = \frac{1}{L}  \sum_{i=1}^L m_i.
\ee
As discussed elsewhere \cite{igloi,faria,igloi1} , this quantity may be experimentally accessible. We now have to 
extrapolate the values obtained for $L \rightarrow \infty$. As expected for aperiodic modulated models, oscillations
occur as depicted in Fig. \ref{fig:mag_oscil}; in order to obtain the value of $m \equiv  m(L \rightarrow \infty)$, we have
used the extrapolation procedure introduced in Ref. \onlinecite{faria}. It simply takes the two last pairs of values
for $m(L)$ and makes a linear extrapolation with each of them. The values $m_1$ and $m_2$ (see Fig.
\ref{fig:mag_oscil}), obtained for $1/L=0$, are then the limits of our estimate for $m$ in the thermodynamic limit.
We then take $m = (m_1 + m_2)/2$ and the error $\Delta m = |m_1 - m_2|/2$. From
Fig.  \ref{fig:mag_oscil} we clearly see that this procedure gives an interval for the magnetization that contains
the true value in the thermodynamic limit, although it overestimates the error. 

   The same procedure was employed to obtain the magnetization for a non-zero magnetic field, which is necessary
to calculate the critical exponents $\delta$ and $\gamma$ (see next section).
  
    \subsection{Critical exponents}
    
        \subsubsection{Critical exponent $\beta$}
    
     Our first attempt to estimate the critical exponent $\beta$ was to fit our data,
obtained in the thermodynamic limit, 
as explained in the previous section, to a log-periodic function:
\be
     m(t) \sim \left( -t \right) ^{\beta} {\cal P} \left[ \log_{10}(-t)  \right],   
\ee
where we assume the following form for the function ${\cal P} \left[ \log _{10}(-t)  \right]$:
\be
  {\cal P} \left[ \log_{10}(-t)  \right] \sim  \left\{ 1 + B \cos \left[ 2 \pi C \log_{10}(-t) + \tau \phi \right]  \right\}.  \label{eq:Pdet}
\ee
Therefore, we obtain for the magnetization:
\be
 m(t) = A \left( -t \right) ^{\beta} \left\{ 1 + B \cos \left[ 2 \pi C \log_{10}(-t) + \tau \phi \right]  \right\}, \label{eq:maglogperiodic}
 \ee
 where $2 \pi$ and $\tau=(\sqrt{5}+1)/2$ are convenient constants for the fitting.
 
    Our  results for $\beta$, using Eq. \ref{eq:maglogperiodic}, are shown in Table \ref{tab:beta}, second column. The amplitude of the
log-periodic term is roughly $5 \times 10^{-3}$ for all values of $r$, except $r=1$ (where this term is not
present). Two results are worth noting: the exponent for $r=1$ (uniform model) is known to be $1/2$; our result,
although near this value
is not consistent with it. Also, the chi-square per degrees of freedom (henceforth referred as $\chi^2$),
is much greater than $1$. This shows that our fitting is not a good
one for the aperiodic models.

\begin{table}[!htp]
	\begin{center}
		\begin{tabular}{|c|c|c|c|}
			\hline  $r$  &   \multicolumn{3}{|c|} {$\beta$} \\
			                     \cline{2-4}
			                     & $(a)$	               &  $(b)$	        & $(c)$	          \\
			\hline $0.5$ & $0.56872(4)$  & $0.5683(2) $   & $0.56824(4)$   \\
			\hline $0.7$ & $0.5439(3)$    & $0.54558(2)$ & $0.545553(6)$ \\
			\hline $1.0$ & $0.489(1)$	      & $0.499 89(4)$& $0.49984(2)$   \\
			\hline $1.3$ & $0.5270(4)$     & $0.53033(5)$& $0.53041(2)$    \\
			\hline $1.5$ & $0.5465(3)$     & $0.54884(4)$& $0.54892(2)$   \\
			\hline
		\end{tabular}
	\end{center}
\caption{Magnetization critical exponent, $\beta$, as a function of the ratio $r$, obtained: $(a)$ fitting
the data to Eq. \ref{eq:maglogperiodic}, $(b)$ using
the logarithmic derivative, and $(c)$ fitting
the data to Eq. \ref{eq:maglogperiodic}, but restricting the interval in $\log_{10}(-t)$. Numbers in parenthesis are uncertainties 
in the last digit.}
\label{tab:beta}
\end{table}

  To improve our estimates for $\beta$, we have made another procedure, which consists in 
calculating the so-called logarithmic derivative, namely:
\be  
    {\cal L}(t) \simeq \frac{d \log_{10}\left[ m(t) \right]}{d \log_{10}(-t)} \sim \beta +  \tilde{B} \cos [ 2\pi C \log_{10}(-t) + \tilde{\phi} ], 
    \label{eq:derlogmag}
\ee
where it is assumed that $B \ll1$ in Eq. \ref{eq:Pdet}. This derivative is numerically obtained and the data
is fitted to the previous equation. Examples of the type of behavior we obtain are depicted in Fig. \ref{fig:mag-t-D_menor_1.0}
for $r=0.7$ ($a$) and $r=1.5$ ($b$). There is a clear oscillation, as predicted by Eq. \ref{eq:derlogmag}; the mean value
of the fitted curve is the exponent $\beta$. Note, however, that for values of $\log_{10}(-t)$ close to $-1$
the behavior departs from the one predicted. Therefore, this interval is not in the scaling region and should not be used to study
the critical behavior. Our fitting is then obtained with the data points in the proper interval. For comparison, we show the
graph o the log-derivative for $r=1$, in Fig. \ref{fig:mag-t-D_1.0}: no oscillation is present but the deviation
from the expected behavior (in this case, a horizontal line) is obtained for $-t$ big enough.
Results for $\beta$ with this procedure are shown in 
Table \ref{tab:beta}, third column. Although the value for $r=1$ does not include the known value for the mean-field approximation, 
it is closer to the expected value than for the previous procedure and correct up to the third 
decimal place. Another improvement with respect to the previous procedure is that the values obtained for $\chi^2$ 
are orders of magnitude smaller: they range from $10^{-3}$ to $10^{-1}$.  The amplitude of the log-periodic term
is small, as expected (the maximum value, for the values of $r$ we studied, is approximately $10^{-2}$, for
$r=0.5$) and increases as we move further away from the uniform case, as expected \cite{andrade1}.

  As a final check, we have made fittings using Eq. \ref{eq:maglogperiodic} but now with a restricted interval of the reduced
temperature $t$. We have used the interval in which the log-derivative behavior is well described
by the data. In Fig. \ref{fig:mag-t-D_menor_1.0} ($a$), for example, this interval is $-4.9 \leq \log_{10}(-t) \leq -2.5$.
The values so obtained of $\beta$ are shown in Table \ref{tab:beta}, forth column: although the result for $r=1$ is closer to
the expected value within the mean-field approximation than for the first fitting procedure, it is not better than the second one. Also, $\chi^2$ has
decreased a great deal, when compared to the first procedure but it is still orders of magnitude greater than
for the log-derivative fitting. Therefore, we will take as our results for $\beta$ those in Table \ref{tab:beta}, third column.

   Finally, we would like to stress the excellent agreement between our results for this exponent and 
those in Ref. \cite{igloi} (see Fig. \ref{fig:beta-Igloi}).

   \subsubsection{Critical exponent $\delta$}

   In order to calculate the exponent $\delta$, one has to study the dependence of the magnetization on the external
uniform magnetic field $h$. As a log-periodic dependence is expected, we also have made all three fitting procedures 
described above for this case. Again the best results are obtained for the second one. 

   More precisely, we assume the dependence of $m$ on $H$ to be (a $\mbox{sgn}(H)$ term is present in the following equation but
   we have omitted it, for clarity):
   \be
         m(H) = A |H|^{ \frac {1} {\delta} } \{ 1 + B \cos [ 2\pi C \log_{10}|H| + \tau \phi ] \}.
   \ee
   Therefore, the logarithmic derivative is given by (again, taken into account that the amplitude of the log-periodic
oscillation is small):
   \be
     {\cal L}(H) \simeq \frac{d \log_{10}(m)}{d \log_{10}|H|} = \frac{ 1 }{ \delta } + \tilde{B} \cos [ 2\pi C \log_{10}|H| + \tilde{\phi} ].
        \label{eq:mlogderH}
   \ee
   The typical behavior is depicted in Fig. \ref{fig:H-D_menor_1.0}: again log-periodic oscillations are present and
the critical exponent $\delta$ is obtained from the previous function (Eq. \ref{eq:mlogderH}).

   The critical exponents are shown in Table \ref{tab:delta-log}. The mean-field value for $r=1$ is $1/3$; our
numerical evaluations agrees with this result up to the fourth decimal place. For the uniform model,
as expected, no oscillation is present in the logarithmic derivative. Finally, for the values of $r$ quoted in 
Table \ref{tab:delta-log}, $\tilde{B}$ (see Eq. \ref{eq:mlogderH})
 varies from $10^{-3}$ to $10^{-4}$ and increases as we move away from the uniform model. 
 The values of $\chi^2$ (not shown) vary from $10^{-3}$ to $10^{-5}$ for the aperiodic models and
equals $10^{-9}$ for the uniform case. These results are evidence of good fittings.
   
\begin{table}[!htp]
	\begin{center}
		\begin{tabular}{|c|c|}
			\hline  $r$   & $ 1 / \delta $ \\
			\hline $0.5$ & $0.37402(2)$ \\
			\hline $0.7$ & $0.358745(3)$ \\
			\hline $1.0$ & $0.33328(1)$ \\
			\hline $1.3$ & $0.34995(2)$ \\
			\hline $1.5$ & $0.360770(5)$ \\
			\hline 
		\end{tabular}
	\end{center}
\caption{Critical exponent $\delta$ as a function of $r$, for fittings to log-derivative functions (Eq. \ref{eq:mlogderH}). 
Numbers in parenthesis are uncertainties  in the last digit.}
\label{tab:delta-log}
\end{table}
      
    \subsubsection{Critical exponent $\gamma$}
    
    We have calculated the susceptibility $\chi(t)$ using two different methods. First, for each reduced temperature $t$, we
 calculate the magnetization for two different (small) magnetic fields and perform a numerical derivative to obtain
 $\chi(t)$. Alternatively, we can differentiate Eq. \ref{eq:sistema} with respect to $H$ and obtain a system
 of equations with $\chi_i(T), i=1,...,N,$ as the variables. Solving for these, we can calculate the susceptibility 
 $\chi(t) \equiv \sum_i \chi_i(T)/N$. 
 
   For the first method, we used the first two procedures quoted in the previous subsections, namely: fitting the data to the
functions
\be
   \chi(t) = A |t|^{-\gamma} \left\{ 1 + B \cos \left[ 2 \pi C \log_{10}|t| + \tau \phi \right]  \right\}
\ee
and
\be
       {\cal L}(t) = \frac{d \log_{10}[\chi(|t|)]}{d \log_{10}|t|} = -\gamma + \tilde{B} \cos [ 2\pi C \log_{10}|t| + \tilde{\phi} ].
        \label{eq:chilogdert}
\ee
But, contrarily to what happened for the two previous critical
exponents, it was not possible to identify a clear log-periodic oscillation for the log-derivative of $\chi(t)$.
This may be due to the importance of more than one harmonic in the behavior of this function \cite{andrade1}; 
we could not test this hypothesis because our data was not enough to obtain one period of the log-periodic oscillation.

Therefore, for the $\gamma$ critical exponent we have only obtained results from the fitting to a log-periodic
function as in the previous equation. These results, although not as precise as the ones obtained from the log-derivative function,
should not be off of the correct values by more than $0.6 \%$, according to the comparison made for the critical
exponents $\beta$ and $\delta$. Our results are shown in Table \ref{tab:gamma}. The mean-field value for the
critical exponent of the uniform case is $1$; our evaluation is $0.1$ $\%$ off.

\begin{table}[!htp]
	\begin{center}
		\begin{tabular}{|c|c|c|c|}
			\hline $r$ & $\gamma$ & $\gamma_{cal}=\beta(\delta-1) $& $ \Delta \gamma (\%) $ \\
			\hline $0.5$ & $0.9557(8)$  & $0.9511(5)$   & $0.5$ \\ 
			\hline $0.7$ & $0.9812(8)$  & $0.97522(5)$ & $0.6$ \\ 
			\hline $1.0$ & $1.0010(3)$  & $1.0000(2)$   & $0.1$ \\ 
			\hline $1.3$ & $0.994(2)$    & $0.9851(2)$   & $0.9$ \\ 
			\hline $1.5$ & $0.9772(7)$  & $0.9725(1)$   & $0.5$ \\ 
			\hline 
		\end{tabular}
	\end{center}
\caption{Susceptibility critical exponent as function of $r$. $\gamma$ stands for the critical exponent calculated using the
fitting procedure described in the text, $\gamma_{cal}$ stands for the calculation using the equality between
exponents $\gamma$, $\beta$, and $\delta$, and the last column shows the percentage difference between the two
estimates for $\gamma$.}
\label{tab:gamma}
\end{table}

     We have also calculated $\gamma$ using the usual scaling relation $\gamma = \beta (\delta-1)$, which
still holds true for anisotropic models (see next subsection), with $\beta$ and $\delta$ taken from the log-derivative fittings. 
The comparison is in Table \ref{tab:gamma}: note that
the discrepancy is $0.9$ $\%$ for the worst case, which confirms our evaluation
that the values would not be off by much more than $0.6$ $\%$.

    As mentioned earlier, another possible method to obtain the susceptibility is to perform a field-derivative of the 
system of equations for the 
magnetization (Eqs. \ref{eq:sistema}), in order to obtain a system of equations
for $\chi_i$. These will be given by the solution of this system, in the same manner that we did for the magnetization.
The results are the same as for the previous method, as expected. In particular, we were not able to characterize
the log-periodic oscillations, either.
   
    \subsubsection{Other critical exponents}
    
    We now turn to the calculation of other critical exponents, using the scaling relation for the free energy 
for anisotropic systems. Due to the presence of the aperiodicity in one dimension, we expect different correlation
lengths in the direction of the aperiodic modulation, $\xi_{\para} \sim t^{\nu_{\para}}$, and along the other directions, 
$\xi_{\perp} \sim t^{\nu}$, with $q \equiv \nu_{\para}/\nu \neq 1$ \cite{binder}.
Assuming the scaling ``ansatz'' for a system in $d$ dimensions
(see Ref. \onlinecite{berche1}, where the scaling relation is proposed for two-dimensional models):
\be
   f_s(t,h,L) = b^{-(d-1+q)} f_s(b^{1/\nu},b^{y_h},L/b),
\ee
where $f_s$ is the singular part of the free energy, $b$ is the rescaling factor, $y_h$ is a scaling exponent,
and $L$ is the linear size of the lattice.

    From the above equation, one can show, in the usual way, the following relations between critical exponents
\cite{stanley}:
\be 
  \gamma = \beta (\delta-1); \;\; \alpha+2\beta+\gamma = 2; \;\; \alpha = 2 - \nu (d-1) - \nu_ {\para}. \label{eq:relacaoexpoentes}
\ee
    
      Therefore, assuming $\nu=1/2$ (since the aperiodic sequence we study is a marginal one \cite{igloi}) 
and $d=4$, the exponents $\alpha$ and $\nu_ {\para}$ assume the values
shown in Table \ref{tab:outros_expoentes}. Note the good accordance with the mean-field values for the uniform model  ($r=1$)
and the expected increase of $\nu_ {\para}$ and decrease in $\alpha$ when we move away from $r=1$.

     Assuming a similar scaling form for the two-point correlation function $\Gamma(x,y)$, where $x$ is the
distance along the aperiodic direction and $y$ is the distance along the remaining $(d-1)$ directions:
\be
     \Gamma(x,y,t) \simeq t^{2 \beta} {\cal G} (x/|t|^{-\nu_{\para}},y/|t|^{-\nu}),  \label{eq:scalingfunction}
\ee
one can show that:
\be
      d-2+\eta_{\para} = 2 \beta/\nu_{\para}; \;\; d-2+\eta = 2 \beta/\nu,  \label{eq:etarelations}
\ee
where $\Gamma(x,0,0) \sim x^{d-2+\eta_{\para}}$ and $\Gamma(0,y,0) \sim y^{d-2+\eta}$.  Therefore, 
the exponent $\eta$ assumes the usual mean-field value, namely $\eta=0$. The values obtained for the
exponent along the aperiodic direction, $\eta_{\para}$, are shown in Table \ref{tab:outros_expoentes},
assuming $d=4$, as before. As expected,
the value for the uniform model is consistent with the known value for the mean-field approximation. However, note that the value for $r=0.5$ is
closer to the uniform results than for $r=0.7$. Since $\eta_{\para}$ is close to zero and it is obtained from $\nu_{\para}$, which itself is calculated from scaling relations, one expect a higher inaccuracy.

  \begin{table}[!htp]
     \begin{center}
     \begin{tabular}{|c|c|c|c|c|c|}
	\hline $r$			    &   $0.5$             &	$0.7$		&	$1.0$	&	$1.3$	&	$1.5$         \\
	\hline $\alpha$		    & $-0.0877(9)$  & $-0.06638(9)$	& $0.0002(2)$  & $-0.0458(3)$	& $-0.0701(2)$  \\
	\hline $\nu_{\para}$     & $0.5877(9)$    & $0.56638(9)$	& $0.4998(2)$	& $0.5458(3)$	& $0.5701(2)$	 \\
	\hline $\eta_ {\para}$   &  $-0.066(4)$    &  $-0.0734(4)$    &  $0.0004(9)$ &  $-0.057(2)$  &  $-0.0746(8)$      \\
	\hline
		\end{tabular}
	\end{center}
\caption{Critical exponents calculated from scaling relations for anisotropic systems.}
\label{tab:outros_expoentes}
\end{table}

\section{Conclusions} \label{sec:conclusion}

   We employ a mean-field approximation to treat an Ising model with aperiodic modulation in one 
spatial direction. The particular aperiodic sequence we use is a marginal one, in the context of the 
Harris-Luck criterion. We calculate many equilibrium critical exponents, including $\nu_{\para}$ and 
$\eta_ {\para}$, assuming $d=4$ to be the upper critical dimension of the model and a particular scaling form for the
singular part of the free energy per site and for the two-point correlation function, suitable for anisotropic models. 
As expected, the exponents (with the exception of $\nu$ and $\eta$) depend on the ratio $r = J_B/J_A$ but obey 
the usual scaling relations for anisotropic models, whenever
possible to test these relations. Our results are in accordance with the known values for the mean-field procedure (uniform
model, $r=1$) or with previous results for the exponent $\beta$ and critical temperatures \cite{igloi}.

\begin{acknowledgments}
The authors would like to thank the Brazilian agencies
FAPESC, CNPq, and CAPES for partial financial support. 
\end{acknowledgments}

\bibliography{references}

\newpage

%%%%%%%%%%%%%%%%%%%%%%%%%%%%%%%
%%% FIG 1
%%%%%%%%%%%%%%%%%%%%%%%%%%%%%%%
\begin{figure}
\begin{center}
%eavevmode
%\epsffile{Figuras/Fibonacci.eps}
\caption{Example of a lattice with an aperiodic modulation given by the Fibonacci sequence.
In the horizontal direction the exchange interactions follow this sequence while in the vertical ``planes''
they assume the same value, equal to the one in the following horizontal bonds. Traced (full) lines
represent $J_A$ ($J_B$) interactions.}
\label{fig:rede} %fig1
\end{center}
\end{figure}

%%%%%%%%%%%%%%%%%%%%%%%%%%%%%%%
%%% FIG 2
%%%%%%%%%%%%%%%%%%%%%%%%%%%%%%%
\begin{figure}
\begin{center}
%\leavevmode
%\epsffile{Figuras/TcT0-Igloi.eps}
\caption{Comparison between our values for $T_c$ (traced line) and those obtained in Ref. \onlinecite{igloi}
(continuous line).
In the text we define the quantity $T_c^0$.}
\label{fig:comparacaoPalagyi} %fig2
\end{center}
\end{figure}

%%%%%%%%%%%%%%%%%%%%%%%%%%%%%%%
%%% FIG 3
%%%%%%%%%%%%%%%%%%%%%%%%%%%%%%%
\begin{figure}
\begin{center}
%\leavevmode
%\epsffile{Figuras/r_1.50-m-L-fit.eps}
\caption{Typical behavior for the magnetization, for fixed reduced temperature $t$ or fixed reduced magnetic field $h$,
as function of the linear size of the lattice, $L$. Note the oscillatory convergence to the thermodynamic limit,
as expected for aperiodic modulated models.}
\label{fig:mag_oscil} %fig3
\end{center}
\end{figure}

%%%%%%%%%%%%%%%%%%%%%%%%%%%%%%%
%%% FIG 4 a and b
%%%%%%%%%%%%%%%%%%%%%%%%%%%%%%%
\begin{figure}[!htbp] 
%		\includegraphics[width=0.49\textwidth]{figuras/mag-t-D_0.70.eps}
%	\hfill
%		\includegraphics[width=0.49\textwidth]{figuras/mag-t-D_1.50.eps}
		\caption{Logarithmic derivative of the magnetization for $r = 0.7$ ($a$) and $r = 1.5$ ($b$). The continuous
		lines are fittings using Eq. \ref{eq:derlogmag} while the points are our numerical data.}
\label{fig:mag-t-D_menor_1.0} %fig4
\end{figure}

%%%%%%%%%%%%%%%%%%%%%%%%%%%%%%%
%%% FIG 5
%%%%%%%%%%%%%%%%%%%%%%%%%%%%%%%
\begin{figure}[!htbp]
%  	\centering
%   \includegraphics[width=0.49\textwidth]{figuras/mag-t-D_1.00.eps}
	\caption{Logarithmic derivative of the magnetization for $r = 1$. The continuous
		lines are fittings using Eq. \ref{eq:derlogmag} while the points are our numerical data. Error bars are
		approximately the same size of the points.}
\label{fig:mag-t-D_1.0} %fig5
\end{figure}

%%%%%%%%%%%%%%%%%%%%%%%%%%%%%%%
%%% FIG 6
%%%%%%%%%%%%%%%%%%%%%%%%%%%%%%%
   \begin{figure}[!htp]
%	\centering
%	\includegraphics[scale=0.6]{figuras/beta-Igloi.eps}
	\caption{Exponent $\beta$ as a function of $r$ from the data obtained in this work (continuous line) and
from the results from Ref. \cite{igloi} (traced line).}
\label{fig:beta-Igloi} %fig6
\end{figure}

%%%%%%%%%%%%%%%%%%%%%%%%%%%%%%%
%%% FIG 7 a and b
%%%%%%%%%%%%%%%%%%%%%%%%%%%%%%%
\begin{figure}[!htbp]
%		\includegraphics[width=0.49\textwidth]{figuras/H-D_0.70.eps}
%	\hfill
%		\includegraphics[width=0.49\textwidth]{figuras/H-D_1.50.eps}
		\caption{Field logarithmic derivative of the magnetization, for $r = 0.7$ ($a$) and $r = 1.5$ ($b$).
The continuous lines are fittings using Eq. \ref{eq:mlogderH} while the points are our numerical data. Error bars are
approximately the same size of the points}
\label{fig:H-D_menor_1.0} %fig7
\end{figure}

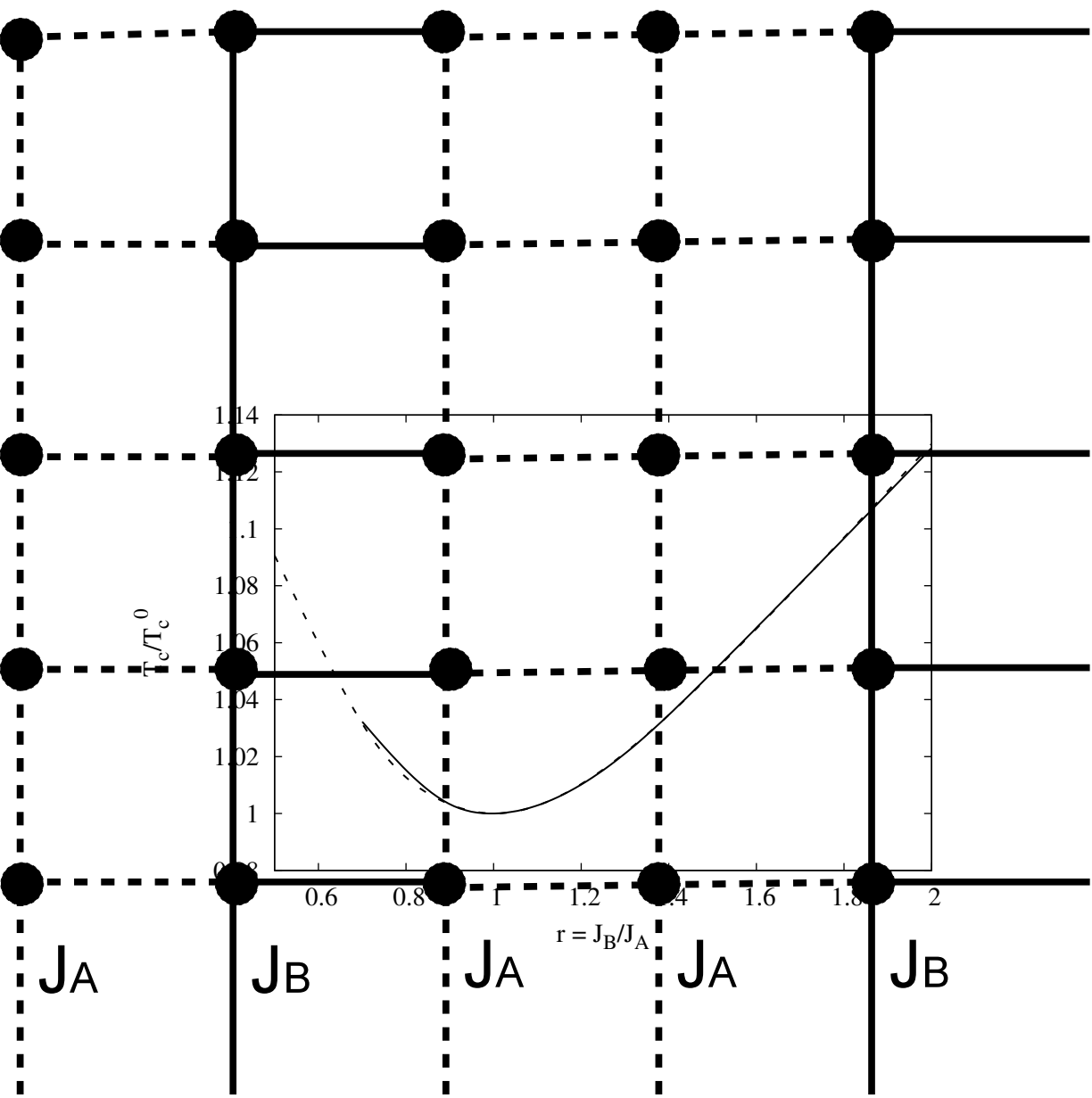
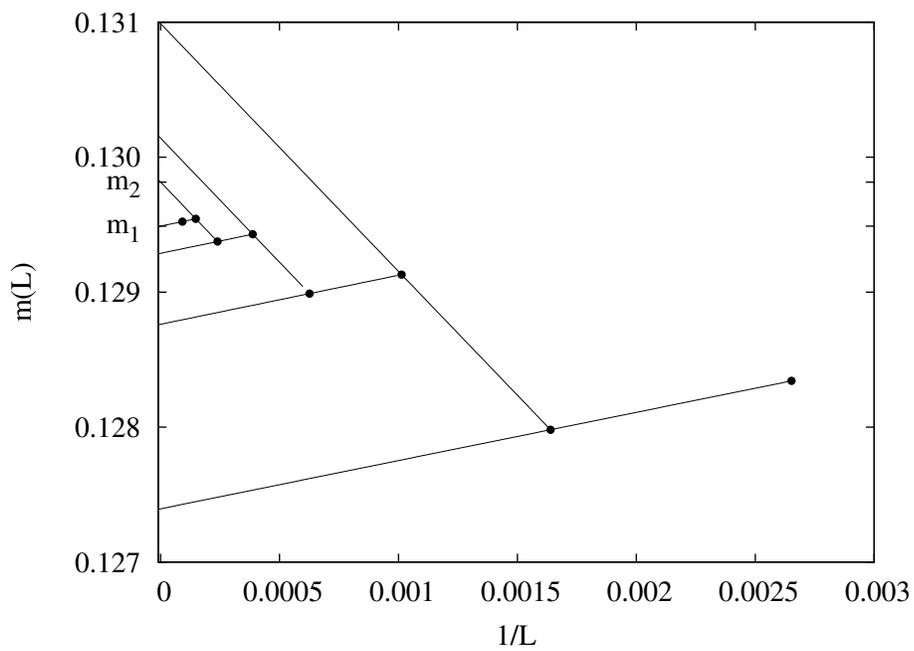
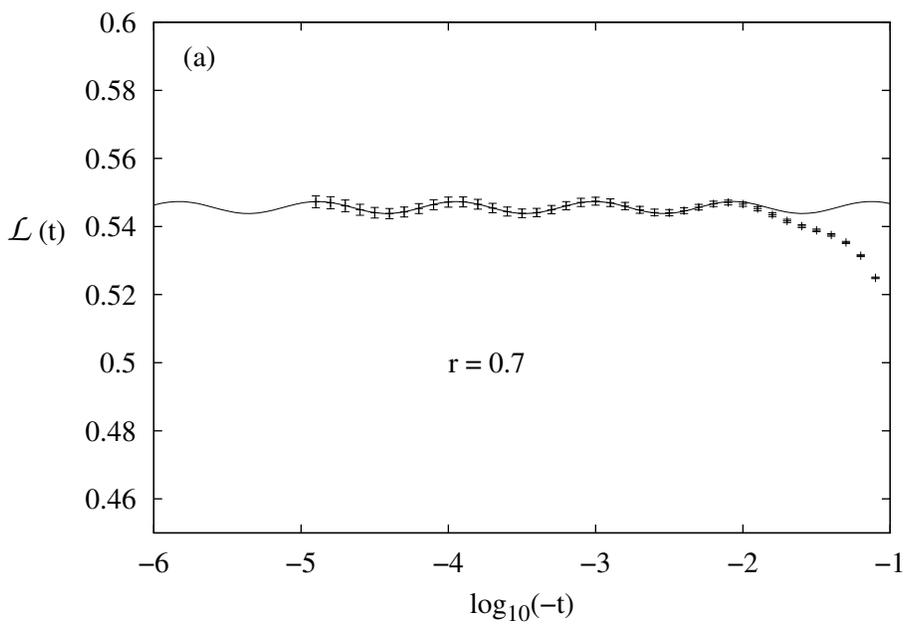
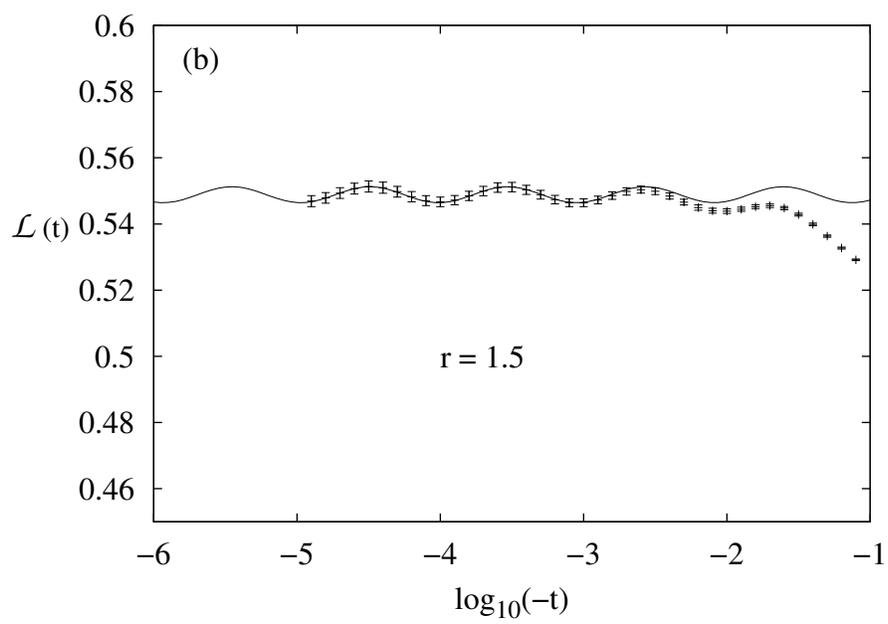
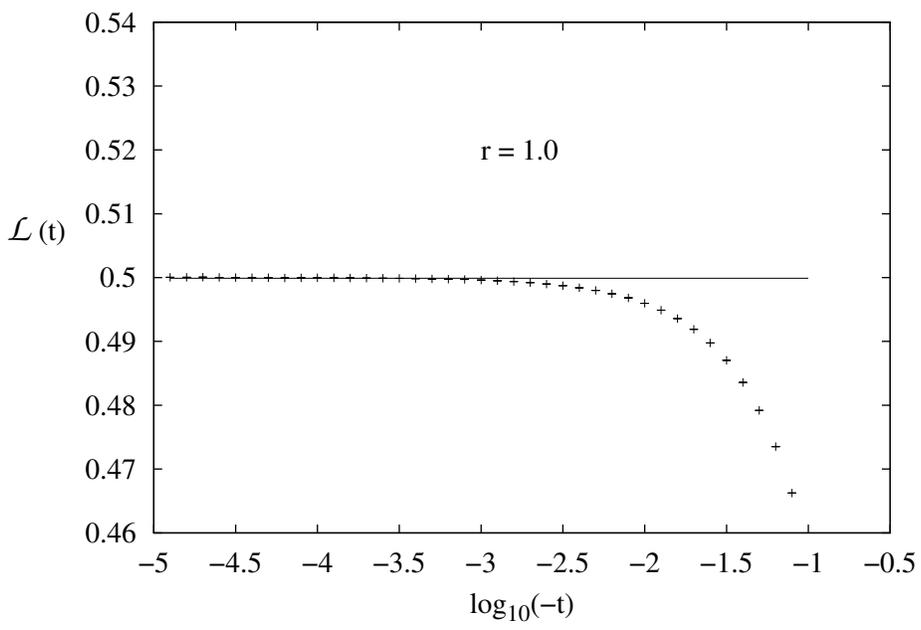
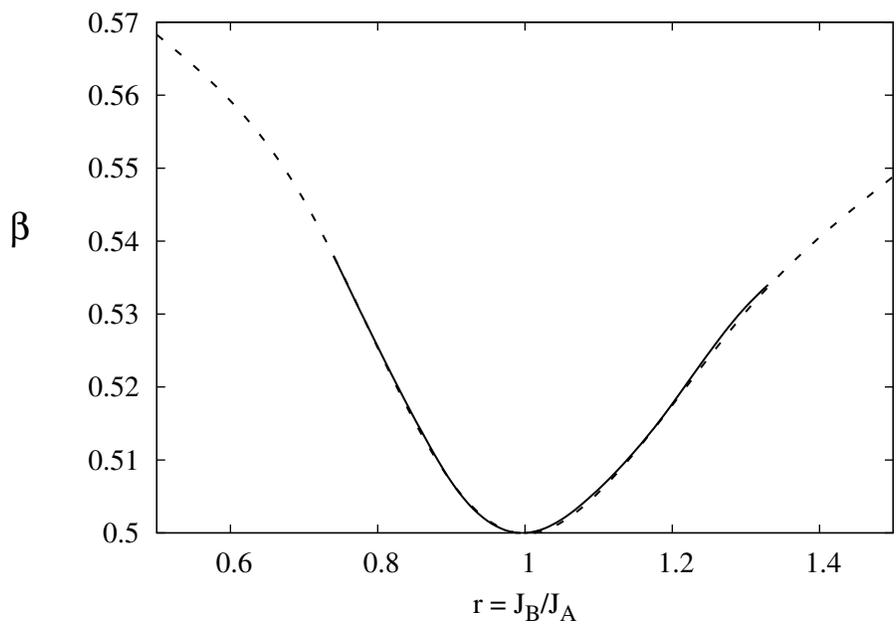
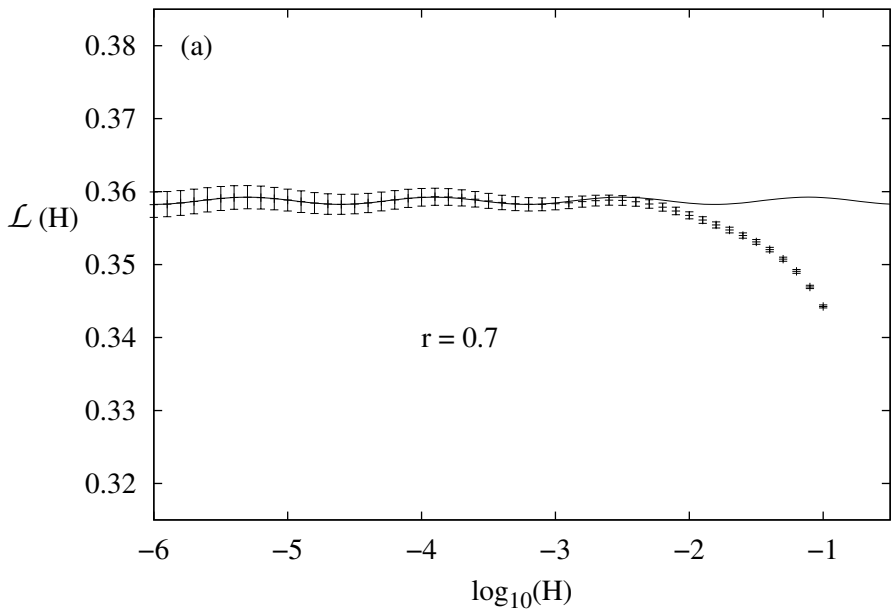
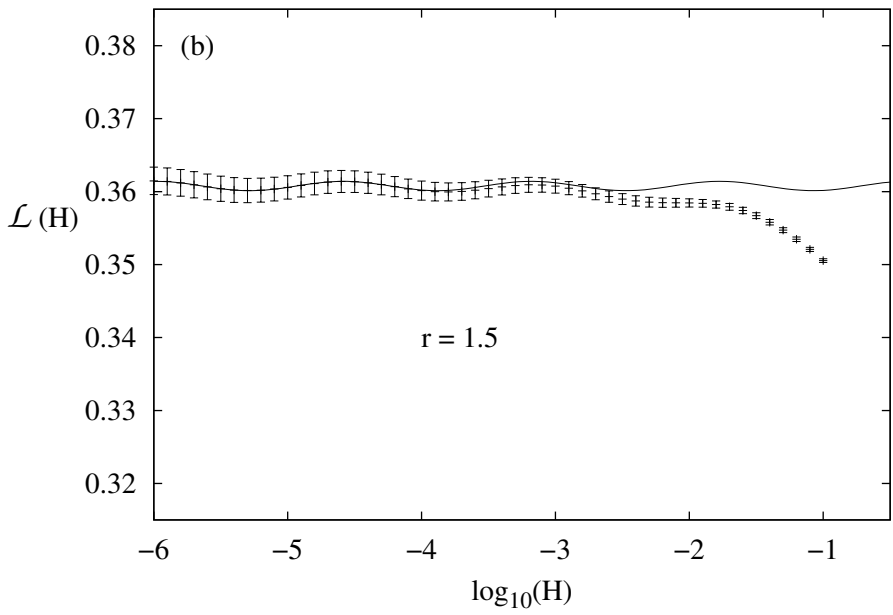
\end{document}